# Precise Neutron Electric Form Factor from Lattice QCD[a]

WALTER WILCOX

*Department of Physics, Baylor University, Waco, TX 76798-7316*

A new calculation of the neutron electric form factor is described. Preliminary results for the connected part on a $20^3 \times 32$ lattice are presented. The methods for calculating the disconnected part are briefly described.

## 1  Introduction

Previous lattice calculations of nucleon electromagnetic form factors on $16^3 \times 24$ and $24 \times 12 \times 12 \times 24$ lattices[1,2] have had decent signals for the proton electric and magnetic form facors, $G^p_e(q^2)$ and $G^p_m(q^2)$, as well as the neutron magnetic form factor, $G^n_m(q^2)$, but very bad signals for the neutron electric form factor, $G^n_e(q^2)$. The most that can be said is that the lattice results agree in sign and magnitude with the expected experimental results. If we are to move beyond the qualitative stage in comparison of lattice data to experiment, this situation must be improved, especially considering the new CEBAF experiments planned[3]. A new, more precise, lattice calculation of $G^n_e(q^2)$ is described here. The lattice used is larger ($20^3 \times 32$), more configurations (50) are employed and better analysis methods are used.

## 2  Simulation Details

Using the usual point nucleon interpolating fields[1], one needs both the two and three-point Green's functions:

$$G_{nn}(t;\vec{p},\Gamma) \equiv \sum_{\vec{x}} e^{-i\vec{p}\cdot\vec{x}} \Gamma_{\alpha'\alpha} \langle \text{vac} | T\left(\chi_\alpha(x)\overline{\chi}_{\alpha'}(0)\right) | \text{vac} \rangle, \qquad (1)$$

$$G_{nJ_\mu n}(t_2,t_1;\vec{q},\Gamma) \equiv -i \sum_{\vec{x}_2,\vec{x}_1} e^{-i\vec{q}\cdot\vec{x}_1} \Gamma_{\alpha'\alpha} \langle \text{vac} | T\left(\chi_\alpha(x_2) J_\mu(x_1) \overline{\chi}_{\alpha'}(0)\right) | \text{vac} \rangle \quad (2)$$

One now forms the connected and disconnected Wick contractions of the proton three-point function. The long Euclidean time limits of these finctions are

$$G_{nn}(t;\vec{p},\Gamma_4) \longrightarrow \frac{E+m}{2E} \frac{|Z|^2 a^6}{(2\kappa)^3} e^{-Et}, \qquad (3)$$

---

[a]Talk presented at the Minneapolis '96 DPF conference.

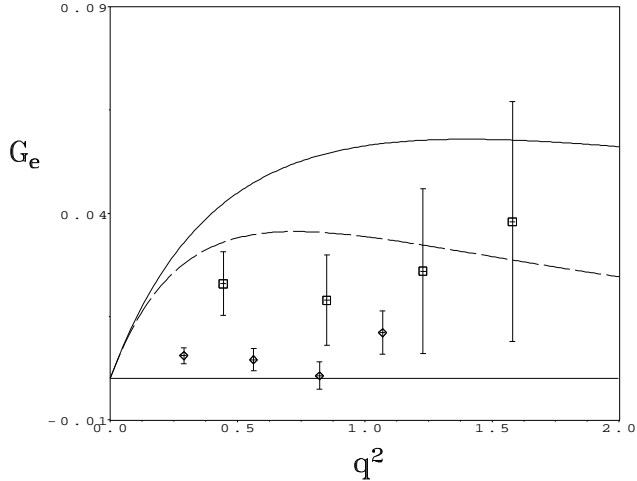

Figure 1: Neutron electric form factor at $\kappa = .152$ as a function of $q^2$ (GeV$^2$).

$$G_{nJ_4n}(t_2, t_1; \vec{q}, \Gamma_4) \longrightarrow \frac{|Z|^2 a^6}{(2\kappa)^3} e^{-m(t_2-t_1)} e^{-Et_1} \left(\frac{E+m}{2E}\right) G_e^n(q^2). \qquad (4)$$

We do not measure these correlation functions directly, but instead analyze the ratio

$$G_e^n(q^2) = \left(\frac{2E}{E+m}\right)^{1-\frac{t_1}{t}} \left(\frac{G_{nJ_4n}(t_2, t_1; \vec{q}, \Gamma_4)}{G_{nn}(t_2; 0, \Gamma_4)}\right) \left(\frac{G_{nn}(t; 0, \Gamma_4)}{G_{nn}(t; \vec{q}, \Gamma_4)}\right)^{\frac{t_1}{t}}, \qquad (5)$$

for $G_e^n$. Fixing $t$ and $t_2$, we look for regions in $t_1$ where the signal stands out from the noise. (The $4 \times 4$ matrix $\Gamma_4 = \frac{1}{2} \begin{pmatrix} I & 0 \\ 0 & 0 \end{pmatrix}$.)

Some results for the connected part of the neutron electric form factor are presented in Fig.1 at $\kappa = .152$ ($\kappa_{cr} = .1564(2)$). In the following we are at $\beta = 6.0$ working on 50 configurations, thermalized by 5000 pseudo-heatbath sweeps and separated by 1000 sweeps. The neutron source is at lattice time site 6 and the zero-momentum sink is at time site 21. The solid and dashed lines in Fig.1 are the "dipole" and "Galster" parameterizations of $G_e^n(q^2)$, respectively, evaluated with the nucleon mass and magnetic moment measured at $\kappa = .152$. The squares represent the previous results for the lowest four nonzero lattice momentums from Ref.1; the diamonds are the new results on the larger lattice. One can see that the error bars are much reduced. Surprisingly, the results are much lower than expected.

One can easily show that the expectation value of the exactly conserved lattice current density, $J_\mu(x) = (\vec{J}(x), i\rho(x)) = i\kappa[\bar{\psi}(x+a_\mu)(1+\gamma_\mu)U_\mu^\dagger(x)\psi(x) - \bar{\psi}(x)(1-\gamma_\mu)U_\mu(x)\psi(x+a_\mu)]$, has a vanishing imaginary part, configuration by configuration. Thus, in forming the disconnected amplitude one is looking at the statistical correlation between the imaginary part of the charge density and the imaginary part of the nucleon two point function, both of which are simply noise by themselves.

Because of the average over gauge configurations, we may replace the point-to-point propagator $M^{-1}(\vec{x}',t';\vec{x},t)$ with a version which is summed over all inital space-time positions in gauge invariant quantities as a consequence of Elitzur's theorem. This method[4] (the "volume method") allows the global measurement of the electromagnetic current, $\sum_{\vec{x}} J_\mu(\vec{x},t)$, necessary for the disconnected part.

The other useful technique for disconnected amplitudes is the $Z(N)$ noise method[5]. Numerically, I have found that $Z(N)$ noise works better than the volume method (the same as $SU(3)$ noise) for local operators such as $\bar{\psi}(x)\psi(x)$, but the volume method is better for the nonlocal $J_\mu(x)$. Since interest here is directed at the disconnected electromagnetic contribution to the nucleon, the volume method has been chosen for the simulations.

## 3 Conclusions, Remarks and Acknowledgements

The method of calculating nucleon form factors on the lattice has been reviewed and a new calculation has been described. An excellent signal is produced for the connected $G_e^n(q^2)$ with error bars significantly smaller than previously. Although the chiral extrapolation has not been done, one can already conclude that the connected results are much smaller than expected.

The method used for the disconnected calculation has been explained. One must look at the correlation of the imaginary parts of the nucleon propagator and charge density. The signal will be hard to detect, but must be present if the phenomenological results are to be recovered from QCD.

This research is partially supported by NSF Grant No. PHY-9401068 and NCSA. I thank Phillip Kalmanson for helping verify the results presented here.